\newif\ifhasjmlrclass
\IfFileExists{jmlr.cls}{
  \hasjmlrclasstrue
  \documentclass[wcp]{jmlr}
}{
  \hasjmlrclassfalse
  \documentclass{article}
}

\usepackage{amsmath,amssymb}
\usepackage{graphicx}
\usepackage{longtable}
\usepackage{booktabs}
\usepackage{lineno}
\usepackage{comment}
\usepackage{array}
\usepackage{adjustbox}
\usepackage{dblfloatfix}
\usepackage{placeins}
\usepackage{pgfplots}
\usepackage{xcolor}
\usepgfplotslibrary{statistics}
\usetikzlibrary{patterns}
\usepackage{float}
\usepackage[most]{tcolorbox}
\usepackage{hyperref}
\hypersetup{pageanchor=false,hypertexnames=false}
\usepackage{url}
\usepackage{natbib}
\usepackage{enumitem}
\pgfplotsset{compat=1.18}
\newcolumntype{P}[1]{>{\raggedright\arraybackslash}p{#1}}
\makeatletter
\@ifundefined{@org@Ginclude@graphics}{}{\let\Ginclude@graphics\@org@Ginclude@graphics}
\let\origincludegraphics\includegraphics
\renewcommand{\includegraphics}[2][]{%
  \IfFileExists{#2}{%
    \origincludegraphics[#1]{#2}%
  }{%
    \fbox{%
      \begin{minipage}[c][0.18\textheight][c]{0.9\linewidth}
      \centering
      \textbf{Missing image}\\
      \texttt{#2}
      \end{minipage}%
    }%
  }%
}
\makeatother
\ifhasjmlrclass\else
\newcommand{\jmlrvolume}[1]{}
\newcommand{\jmlryear}[1]{}
\newcommand{\jmlrworkshop}[1]{}
\newcommand{\editors}[1]{}
\newcommand{\Name}[1]{#1}
\newcommand{\Email}[1]{\texttt{#1}}
\newcommand{\addr}{\\}
\newcommand{\AND}{\and}
\newenvironment{keywords}{\par\noindent\textbf{Keywords: }}{\par}
\makeatletter
\renewcommand{\title}[2][]{\gdef\@title{#2}}
\providecommand{\@jmlrpages}{}
\makeatother
\fi
\jmlrvolume{353,}
\jmlryear{2026}
\jmlrworkshop{ACML 2026}
\title[GRIDEX]{GRIDEX: Region-Grounded Forensic Explanations for Vocoder-Based Synthetic Speech}
\newif\ifanonsubmission
\anonsubmissionfalse
\ifanonsubmission
\author{%
\Name{Anonymous Authors}\\
\addr Anonymous Institution
}
\else
\author{%
\Name{Thi Ngan Ha Do} \Email{do.nganha11@gmail.com}\\
\Name{Tingmin Wu}\thanks{Corresponding author.} \Email{Tina.Wu@csiro.au}\\
\Name{Alsharif Abuadbba} \Email{Sharif.Abuadbba@csiro.au}\\
\Name{Kristen Moore} \Email{Kristen.Moore@csiro.au}\\
\addr CSIRO, Australia
}
\fi
\editors{Andy Song, Bo Han and Sarah Erfani}

\begin{document}
\makeatletter
\let \@jmlrpages \@empty
\makeatother
\maketitle
\begin{abstract}
The advancement of speech generation technologies has made artificial speech increasingly realistic. Although modern classifiers achieve high accuracy in deepfake detection, they do not indicate where spoof cues appear in the spectrogram and what they imply acoustically, limiting their usefulness in forensic settings. Manual analysis of full spectrograms is resource-intensive, so evidence should narrow attention to the most diagnostic regions. Moreover, existing explainability methods have limited capabilities in connecting contextual attributes to localized evidence, making explanations harder to verify. To overcome this limitation, we propose \textsc{GRIDEX}, a pipeline that, given a detector-flagged spectrogram, generates forensic explanations of its anomalies. The pipeline (i) selects top-$K$ anomalous regions in the spectrogram and (ii) produces an explanation for each anomaly. The explanations follow a schema of categorical acoustic fields, including temporal, spectral, and phonetic information, as well as interpretation text. To our knowledge, this is the first framework to generate structured forensic explanations using region-grounded evidence in a controlled vocoder-based setting. \textsc{GRIDEX} is trained with a staged learning recipe that combines supervised fine-tuning (SFT) with Group Relative Policy Optimization (GRPO). Experiments on our dataset show improved artifact localization and explanation grounding quality over strong vision-language model (VLM) baselines. %

\end{abstract}
\section{Introduction}
Deepfake speech increasingly threatens trust in voice-based systems and audio evidence \citep{zhang2025audio}, making it harder for users and systems to distinguish genuine voices from synthetic ones. There is a growing need to understand not only whether audio is spoofed, but also where suspicious evidence appears in a spectrogram and what acoustic features deviate from bona fide speech. Multimedia forensics is the scientific process of isolating anomalies and analyzing the evidence \citep{singh2025unmasking}, going beyond a global authenticity decision. In the audio domain, this corresponds to investigating artifacts that generative models leave in the signal. In practice, these cues are often inspected in a time-frequency representation, which is also the dominant visualization used by audio explanation methods \citep{zhang2025audio}. However, interpreting visual representations, such as spectrograms, is inherently challenging, as spectrograms are technical \citep{zhang2022towards}, and deepfake cues manifest as subtle, distributed textures \citep{coletta2025anomaly}.

Forensic workflows require explanations to be falsifiable, meaning each claim must be traceable to a set of evidence. Audio deepfake forensic analysis, for that reason, requires explanations that are both localized and verifiable, linking each claim to a specific spectrogram region \citep{Ge2022ExplainableDeepfakeSHAP, yi2023add} and its context \citep{gupta24b_interspeech}.
A widely adopted approach is explainable AI (XAI), which provides a means to identify which evidence indicates the synthetic nature of artificially generated speech. Existing audio XAI methods include post-hoc attribution, which produces saliency without semantic interpretation \citep{kasmi2025one} or free-form textual rationales \citep{akman2024audio} that are difficult to audit \citep{lyu2024towards}. While saliency points to areas of interest for the model, it provides limited information about the artifact's nature and its basis for indicating spoofing. On the other hand, free-form rationales can be compelling even in the absence of actual evidence. This poses a potential risk in a forensic scenario, where the rationale could include features not supported by the actual spectrogram content. Neither method jointly optimizes localization and structured explanation under verifiable constraints. These gaps motivate us to address the following research question:
\emph{How can we design an explainable audio forensics system that generates region- and acoustically grounded explanations aligned with spectrogram evidence?}

To this end, we propose GRIDEX, a region-grounded forensic explanation pipeline for vocoder-based audio deepfake spectrogram analysis. Vocoders are core components of many modern speech generation pipelines, making this setting a controlled first step toward grounded deepfake explanation analysis. In comparison to post-hoc saliency or free-form rationales, our pipeline produces explanations that are auditable against the spectrogram. Given a spectrogram, the pipeline operates in two stages. First, it detects a set of the most suspicious regions, prioritizing where artifact evidence appears strongest.
Second, for each detected region, it produces an explanation that describes the region's acoustic context. The output therefore consists of (i) a ranked list of suspicious regions and (ii) a set of per-region explanations, enabling analysts to inspect where the evidence is and what it suggests. Our main contributions are:
\begin{itemize}
    \item We introduce GRIDEX, a framework for region-level localization and structured explanation of vocoder-induced artifacts in synthetic speech spectrograms.
\item We construct a region-grounded explanation dataset from the VocV4 corpus containing 20,512 spectrogram samples and 61,536 structured region-level explanations derived from bona fide and vocoder-generated speech pairs.
\item We design a task-specific staged optimization strategy combining SFT and GRPO to improve artifact localization and structured explanation generation.
\item Experiments on the curated VocV4 dataset show that GRIDEX achieves substantially higher artifact-localization performance than strong VLM baselines and generates structured, region-grounded explanations.
\end{itemize}

\section{Problem Definition}

\subsection{Task Overview}
\label{sec:task_overview}
GRIDEX operates as a post-hoc explanation module for samples flagged by an upstream detector. Given a flagged sample, it explains vocoder-induced anomalies through two objectives: artifact localization and artifact interpretation.
These two objectives are achieved through two sequentially inserted queries into a VLM. These queries will be referred to as Query 1 and Query 2, respectively. The input spectrogram $\mathbf{S}$ is first overlaid with a fixed grid $G \times G$, where $G$ is the number of divisions along each dimension; this grid divides the spectrogram into cells indexed by unique cell identifiers (ID).

\begin{figure}[tbp]
\centering
\includegraphics[width=0.9\linewidth]{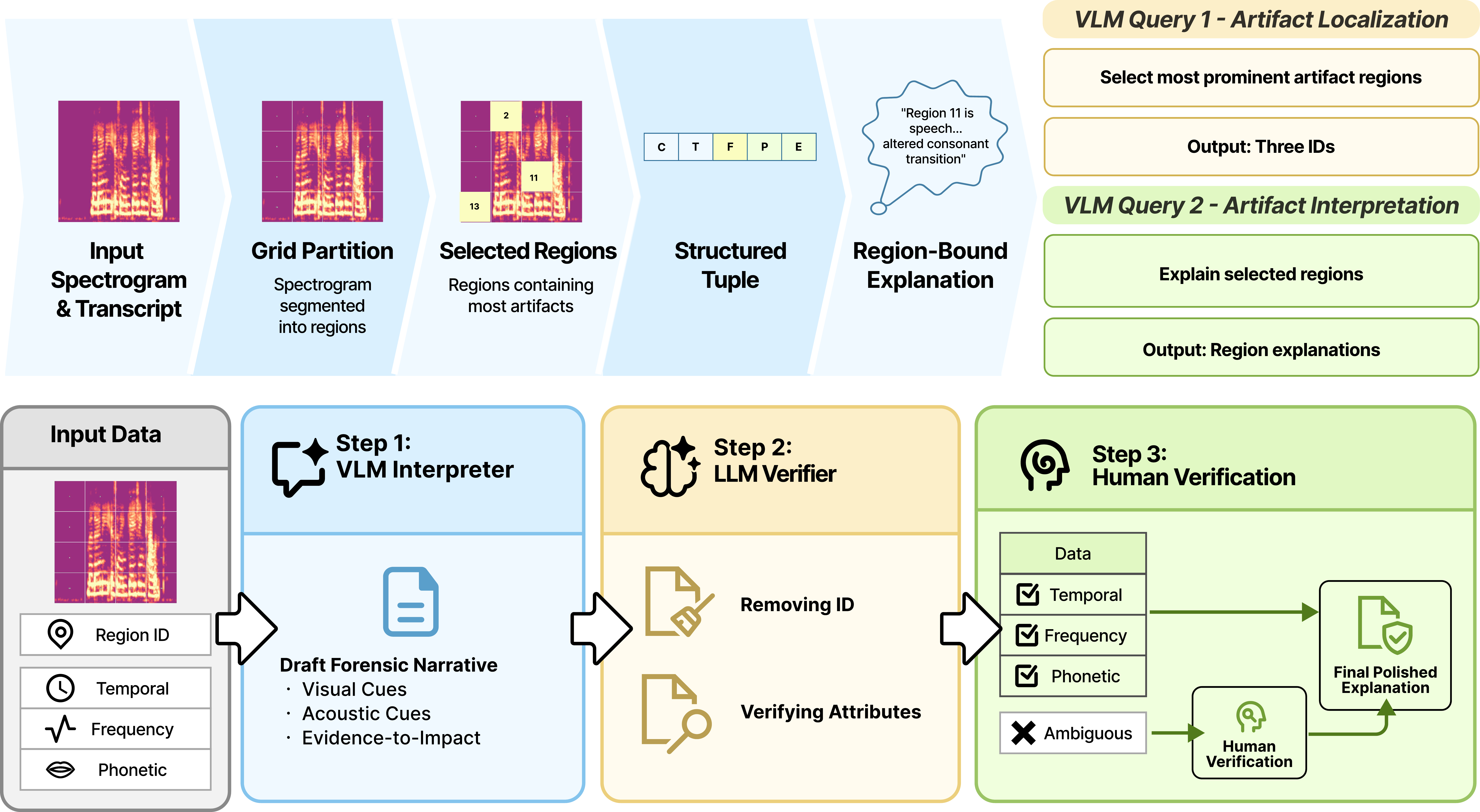}
\caption{Overview of GRIDEX. The top panels define the grid-grounded explanation format and inference procedure, where Query~1 localizes regions and Query~2 explains them. The bottom panel summarizes the dataset construction.}
\label{fig:pipeline}
\end{figure}

\textbf{Query 1.}
The partitioned spectrogram is used as an input for Query~1 to identify where the most suspicious evidence occurs in the spectrogram. The model performs this by selecting a set of cell IDs that exhibit the strongest spoof cues and outputting an ordered list of the top-$K$ cells ranked by artifact evidence, from most to least prominent. In our experiments, we set $K=3$:
\begin{equation*}
\hat{\mathcal{C}} = [\hat{c}_1, \hat{c}_2, \hat{c}_3].
\end{equation*}

\textbf{Query 2.}
Query 2 takes the three ranked cell IDs from Query 1, the partitioned spectrogram, and the corresponding transcript as input. For each selected region, its attributes are extracted and then used to build an explanation of the artifact and its impact on audio authenticity. For each input cell $\hat{c}_i$, in the order of $\hat{\mathcal{C}}$, Query~2 outputs an attribute tuple:
\begin{equation*}
y_i = (\hat{T}_i, \hat{F}_i, \hat{P}_i, \hat{E}_i).
\end{equation*}
$\hat{T}_i$, $\hat{F}_i$, and $\hat{P}_i$ are the predicted temporal, frequency, and phonetic attributes, and $\hat{E}_i$ is the evidence statement for cell $\hat{c}_i$.

\subsection{Why Structured Explanations?}
In forensic settings, a diagnosis is most reliable when supported by evidence. Current explainability methods often provide limited forensic value because their claims are weakly grounded in or unsupported by spectrotemporal evidence \citep{grinberg2025data,chou2026iclad}. To address this, our framework provides structured descriptions that (i) are verifiable with visible cues and (ii) are interpreted using a standard set of acoustic properties. This representation allows for swift inspection, particularly in the analysis of a large case volume, in addition to providing a narrowed list of candidate regions. Using a predetermined schema allows the structured rationales to resemble an evidence record, similar to how forensic data is documented and analyzed. Additionally, the annotations at the field level allow for the aggregation of data, thereby characterizing patterns of artifact types. Each explanation therefore includes the categorical fields $(T,F,P)$ alongside the evidence statement.

\section{Data Construction and Region Annotation}
\label{sec:data_construction}
Most prior deepfake audio datasets offer spoof labels, providing limited value for interpretability research. We therefore construct a dataset that provides region-level artifact supervision, structured metadata, and forensic explanations for each spectrogram. %

\subsection{Localization Supervision}
\label{sec:localization_supervision}

\subsubsection{Difference Masks}
\label{sec:diffmap_grid}

Supervised artifact localization requires targets for where spoof deviations occur. We derive these targets from parallel bona fide--spoof audio pairs produced by self-vocoding, where speaker identity and linguistic content are aligned. Therefore, residual differences primarily reflect spoof artifacts. In contrast, such residual supervision is less identifiable for output from full text-to-speech (TTS) or voice-conversion (VC) pipelines, where timing, prosody, or content may vary. These residual-derived targets provide computational localization supervision by identifying deviations introduced during self-vocoding. However, residual magnitude alone does not establish perceptual significance.

Following \citet{grinberg2025data}, we compute a difference mask from Short-Time Fourier Transform (STFT) magnitude spectrograms $M_b$ and $M_s$, where $M_b$ and $M_s$ denote the magnitude spectrograms of the bona fide and spoofed audio, respectively. We apply 2-D Gaussian smoothing $G(\cdot)$ (kernel $(3,11)$; variances $(3,5)$) to suppress isolated noise, and compute the normalized absolute difference at time frame $t$ and frequency bin $f$, with $\epsilon=10^{-8}$:

\begin{equation*}
D(t,f)=\frac{\left|G(M_s)(t,f)-G(M_b)(t,f)\right|}{G(M_b)(t,f)+\epsilon}.
\end{equation*}
We binarize $D$ by marking values above the 95th percentile of its finite values as 1 and all others as 0, forming the difference mask. Each region is scored by summing the mask values, and regions are ranked to define the top-$K$ supervision for Query 1. %

\subsubsection{Spectrogram Segmentation Methods}
\label{sec:multi_granularity}

\begin{figure}[tbp]
\centering
\includegraphics[width=0.8\columnwidth]{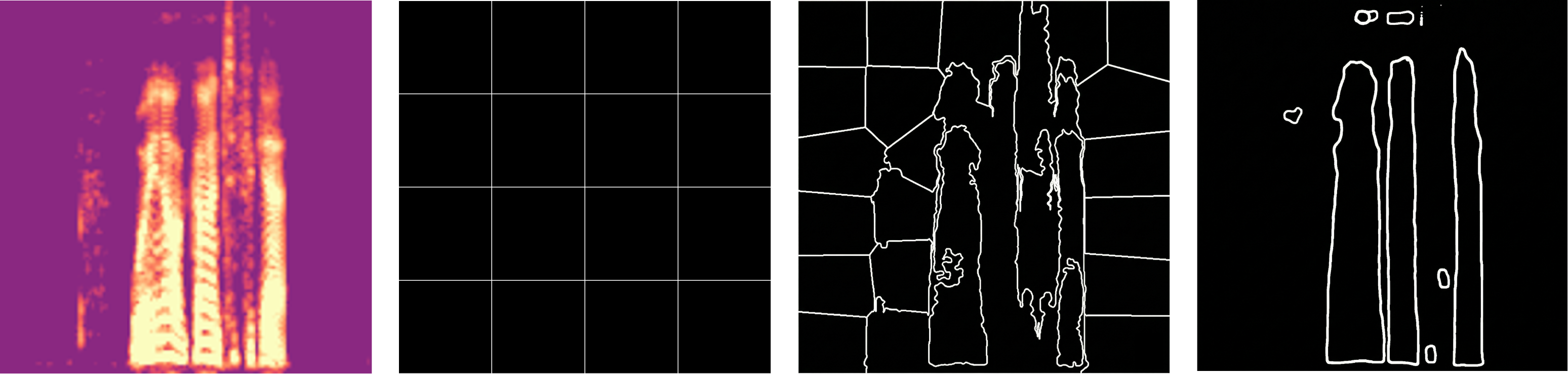}
\caption{Spectrogram segmentation methods. From left to right: raw spectrogram, uniform grid, SLIC superpixels, and SAM masks.}
\label{fig:segmentation_methods}
\end{figure}

We evaluate three segmentation strategies \citep{Zhang2024GPT4VAD}: (1) uniform grids, (2) Simple Linear Iterative Clustering (SLIC) superpixels \citep{achanta2012slic}, and (3) Segment Anything Model (SAM) \citep{Kirillov2023SegmentAnything}. Each strategy produces a set of artifact regions by overlaying its division masks onto the difference masks (Sec.~\ref{sec:diffmap_grid}). As illustrated in Fig.~\ref{fig:segmentation_methods}, SAM often produces texture-like spectrogram masks consisting of isolated points and overly large regions. Consequently, supervision becomes less informative, since coarse regions tend to dominate localization, whereas finer regions provide limited signal. We therefore focus on grids and superpixels, and select the granularity whose top-$K$ regions jointly cover at least $50\%$ of the total difference evidence. In our experiments, we set $K=3$. Both grid ($4{\times}4$) and SLIC (40 target superpixels, 20 compactness weight) satisfy this threshold. We adopt the $4{\times}4$ grid in the main pipeline, as it yields higher localization performance than superpixels under the same training recipe (Sec.~\ref{sec:ablation_granularity}). %

\subsection{Explanation Supervision}
\label{sec:explanation_supervision}

\subsubsection{Region Metadata}
\label{sec:region_metadata}

\begingroup\emergencystretch=2em
The fields take the following categorical values: $T\in\{\texttt{speech},\texttt{non-speech}\}$, $F\in\{\texttt{low},\texttt{mid},\texttt{high}\}$, and $P\in\{\texttt{vowel},\texttt{consonant},\texttt{none}\}$. Independently constructed references $(T,F,P)$ are derived from $\mathrm{Mask}(t,f)$ and Montreal Forced Aligner (MFA) phone alignments. $T$ is assigned by overlap with MFA phone intervals, $F$ by quantizing the mean vertical coordinate of active mask pixels into three frequency bands, and $P$ by the maximum overlap phone.
\par\endgroup

\subsubsection{Automatic Region Annotation Pipeline}
\label{sec:annotation_pipelines}
Our explanation schema requires an evidence statement $E_i$ that describes the anomaly and its probable audio impact. There are limited resources available for such rationales, and manual annotation at this granularity is costly. To overcome this, we generate $E_i$ via a two-step automatic pipeline \citep{xie2026interpretable} and manually verify mismatched samples (Fig.~\ref{fig:pipeline}).

\textbf{Step 1: VLM Interpreter.}
\begingroup\emergencystretch=2em
For pseudo-reference construction, the VLM input includes the segmented spectrogram, the selected region $c_i$, its structured attributes $(T_i,F_i,P_i)$, and reference cues. These cues combine selected forensic-phonetic categories from the Phonetic Assessment of Spoofed Speech (PASS) framework \citep{Lee2023PASSIAFPA} with common vocoder artifact patterns, and are used only during annotation construction to regularize $E_i$ and reduce unconstrained free-form rationales.
\par\endgroup
Given these inputs, Qwen3-VL-30B-Thinking produces a reasoning chain with three components: (i) it compares the phonetic and time-frequency context against the expected structure of bona fide speech, (ii) it describes the visual distortion in shape or texture, and (iii) it writes a short forensic narrative linking the visual cue to a possible acoustic implication. We retain this narrative as $E_i$.

\textbf{Step 2: Large Language Model (LLM) Verifier.}  
Qwen3-235B-Instruct refines the draft, removes region identifiers, and verifies consistency with metadata fields ($T_i$, $F_i$, $P_i$). Uncertain attributes are marked as \texttt{ambiguous} and manually updated.

\subsection{Dataset Quality}
\subsubsection{Vocoder-wise stability of top-3 regions}
\label{sec:quality_vocoder_stability}

To inform localization, the per-vocoder top-3 derived from the difference masks must capture stable region-ID sets, meaning that supervision reflects artifact patterns introduced by the vocoders and not random noise. This stability is measured by Jaccard overlap (how often similar artifact regions occur across samples generated by one vocoder). Compared to a random baseline of the expected mean Jaccard (mean $J$ $\approx 0.120$), the similarities between the vocoders are greater. HiFiGAN and WaveGlow show the strongest stability (mean $J$=0.428 and 0.390), with typical overlap of 2/3 of the regions, whereas Hn-NSF and NSF-HiFiGAN are less stable (mean $J$=0.207 and 0.189), with typical overlap closer to 1/3 of the regions. The within--between gaps are also largest for HiFiGAN ($\Delta J$=0.289) and WaveGlow ($\Delta J$=0.346), indicating stronger vocoder-specific localization patterns.

\subsubsection{Structured metadata quality}
\label{sec:metadata_quality}

We audit the structured region metadata ($T$, $F$, $P$) and the associated $E_i$. Each region is assigned categorical labels using deterministic extraction rules (Sec.~\ref{sec:region_metadata}). The structured metadata is populated by construction: all samples contain exactly three region tuples, and each tuple includes in-vocabulary values with no empty fields.

\textbf{Parse success and verifier agreement.} We measure whether the extracted fields from $E_i$ are consistent with the reference fields ($T$, $F$, $P$). Over the full dataset of 61{,}536 samples, 51{,}278 (83.3\%) match all three fields after LLM verification, while 7{,}103 (11.5\%) match 2/3 fields, 3{,}087 (5.0\%) match 1/3 fields, and 68 (0.11\%) match none of the fields.

\textbf{Human correction.} For samples that do not fully satisfy the verifier checks, mismatches arise from text inconsistencies. Therefore, we manually edit the explanations to align with the fields. After correction, all released $E_i$ satisfy the same verifier checks.

\section{GRIDEX System Design}
\subsection{Overview}
\label{sec:system_challenges}
GRIDEX is a two-step VLM pipeline that first localizes the top-$K$ artifact regions on a spectrogram (Query 1) and then generates structured explanations for those selections (Query 2). This conditioned multi-turn setting introduces two training challenges: errors from the earlier turn propagate into the latter \citep{zhao2025uncertainty} and jointly updating the same parameters across the two turns can overwrite stage-specific behavior \citep{che2025lora}. To address the first issue, Query 2 is refined with a two-phase GRPO schedule (prediction-conditioned and ground-truth (GT)-conditioned refinement). To address the second, we use task-specific parameter-efficient fine-tuning (PEFT) adapters for the two queries. The following subsections describe the staged SFT/GRPO procedure.

\subsection{Supervised Fine-Tuning}
We apply SFT with the standard autoregressive likelihood objective, using different supervised token spans for the two stages. SFT-1 supervises the Query~1 to learn strict list formatting, while SFT-2 supervises the Query~2 to learn the structured tuple schema. In our PEFT setup, the backbone is frozen, and only the stage-specific adapters are updated.

\subsection{Group Relative Policy Optimization}
\label{sec:grpo}

After SFT, GRPO-1 optimizes Query~1, while GRPO-2 refines Query~2 through sequential prediction-conditioned and GT-conditioned phases. Phase~A uses the predicted region IDs of Query~1 to expose Query~2 to inference input, and Phase~B continues from Phase~A with GT IDs to stabilize field accuracy and explanation quality. The reward design follows the constraints of each GRPO stage and addresses potential failure modes. %

\subsubsection{GRPO-1: Region Localization}

\textbf{Hit reward ($H$).}
Let $\hat{C}$ denote the predicted top-3 region-ID set and let $C_{\mathrm{gt}}$ denote the GT top-3 set. To reward retrieving the correct region set regardless of order, we define:
\begin{equation*}
H = \frac{|\hat{C} \cap C_{\mathrm{gt}}|}{3}.
\end{equation*}

\textbf{Ranking reward (\(D\))} is nDCG@3 with graded relevance 3, 2, and 1 for GT ranks 1--3, and 0 otherwise.

\textbf{Format reward (\(B_{\mathrm{fmt}}\)).}
We use a binary indicator that rewards the list schema $[\hat{c}_1,\hat{c}_2,\hat{c}_3]$ with exactly three unique integers in $[1,16]$.

\begin{equation*}
R^{(1)} \;=\; 1.0\,H \;+\; 0.3\,D \;+\; 0.1\,B_{\mathrm{fmt}}.
\end{equation*}

\subsubsection{GRPO-2: Region Explanations}

Let the model output $\{y_i\}_{i=1}^{3}$, where $y_i=(\hat{T}_i,\hat{F}_i,\hat{P}_i,\hat{E}_i)$ corresponds by position to input region $\hat{c}_i$, and let the GT set be $\{(c_j,T_j,F_j,P_j,E_j)\}_{j=1}^{3}$. For $\hat{c}_i \in C_{\mathrm{gt}}$, reference-dependent rewards use the GT tuple with the same region ID, not the same output position. We define four per-tuple reward terms.

\textbf{Text reward (\(R_i\)).}
When $\hat{c}_i \in C_{\mathrm{gt}}$, we compute $\mathrm{ROUGE\text{-}L}$ between the generated evidence text $\hat{E}_i$ and the matched GT text $E_{j(i)}$.

\textbf{Field reward (\(A_i\)).}
When $\hat{c}_i \in C_{\mathrm{gt}}$, we measure exact-match accuracy against the matched GT fields:
\[
A_i=\frac{1}{3}\Bigl(\mathbb{I}[\hat{T}_i=T_{j(i)}]+\mathbb{I}[\hat{F}_i=F_{j(i)}]+\mathbb{I}[\hat{P}_i=P_{j(i)}]\Bigr).
\]

\textbf{Consistency reward (\(C_i\)).}
Although we use an LLM for field extraction in other sections, we use a rule-based lexical parser aligned with the field vocabulary to extract $(\tilde{T}_i,\tilde{F}_i,\tilde{P}_i)$ from $\hat{E}_i$, and score agreement with the declared fields:
\[
C_i=\frac{1}{3}\Bigl(\mathbb{I}[\hat{T}_i=\tilde{T}_i]+\mathbb{I}[\hat{F}_i=\tilde{F}_i]+\mathbb{I}[\hat{P}_i=\tilde{P}_i]\Bigr).
\]

\textbf{Format reward (\(B_i^{\mathrm{fmt}}\)).}
A binary indicator that equals 1 if tuple $i$ parses under the required schema with valid labels and a non-empty $\hat{E}_i$.

\textbf{Gating.}
Let $m_i\in\{0,1\}$ indicate whether $\hat{c}_i\in C_{\mathrm{gt}}$. When $m_i=0$, we mask out the text and field terms. The completion must parse into exactly three tuples; otherwise $R^{(2)}=0$:
\begin{equation*}
R^{(2)}=\tfrac13\sum_{i=1}^3\!\left(m_i\!\left(R_i+0.5A_i\right)+0.1C_i+0.1B_i^{\mathrm{fmt}}\right).
\end{equation*}

\subsection{Implementation Details}
\label{sec:impl_details}

Backbone: Qwen2.5-VL-3B-Instruct. Inputs: Query~1 receives a grid-segmented spectrogram. Query~2 receives the same image, transcript, and selected region IDs. Training: We use task-specific low-rank adaptation (LoRA) adapters for the two queries: Query~1 is trained with SFT-1 and GRPO-1, while Query~2 is trained with SFT-2 and the two-phase GRPO-2 curriculum in Sec.~\ref{sec:grpo}. All stages are implemented in MS-Swift. %

\section{Evaluation}
\subsection{Dataset}
We use the VocV4 vocoder dataset~\citep{Wang2022SpoofedTrainingDataNeuralVocoders} to generate reference annotations (Sec.~\ref{sec:data_construction}). 
VocV4 is constructed by re-synthesizing ASVspoof2019 LA bona fide utterances with four neural vocoders: HiFiGAN, NSF-HiFiGAN, Hn-NSF, and WaveGlow. This produces four spoofed versions of each source utterance. We use 10,320 training samples for model training and fine-tuning, and 10,192 development samples for evaluation. These correspond to 2,580 and 2,548 bona fide source utterances \citep{wang2020asvspoof}, respectively. Using the released VocV4 protocols, we verified that the training and evaluation partitions have no overlapping source utterance IDs or speaker IDs, and that all four vocoders are represented in both partitions.

\subsection{Evaluation Metrics}
\label{sec:metrics}
We assess three dimensions: (i) localization accuracy, (ii) structured field correctness, and (iii) explanation grounding quality.

\textbf{Localization.}
For the ordered top-$3$ region list, we report Recall (R@3), nDCG, and mAP, which are standard in object detection and ranking tasks. R@3 captures hit rate, nDCG measures ranking quality, and mAP summarizes ranked precision over the list.

\textbf{Structured Fields.} FieldAcc macro-averages exact-match accuracy of $\hat{T}_i$, $\hat{F}_i$, and $\hat{P}_i$ against reference attributes for input cell $\hat{c}_i$, matched by position.

\textbf{Explanation Grounding.} CovAvg macro-averages coverage of the three field cues extracted from $\hat{E}_i$ by the LLM verifier described in Sec.~\ref{sec:annotation_pipelines}.
Explanation quality uses ROUGE-L and BERTScore F1 (BERTF1) against pseudo-GT rationales.

\subsection{Results}

\subsubsection{Quantitative Analysis}

Table~\ref{tab:baselines_combined} reports the performance of VLM baselines on the localization task, the explanation task, and the end-to-end (E2E) pipeline. All baselines use identical prompts, a $4{\times}4$ grid, and top-3 selection, with other experimental settings held constant.

\begin{table}[H]
\centering
\footnotesize
\setlength{\tabcolsep}{2pt}
\renewcommand{\arraystretch}{1.08}
\caption{Pipeline evaluation for Query~1 and Query~2. Zero-shot (ZS) reports Query~1, Oracle reports Query~2 with Oracle IDs, and E2E chains. The best result for each metric is shown in bold.}
\label{tab:baselines_combined}

\begin{adjustbox}{max width=\columnwidth}
\begin{tabular}{llccc|cccc}
\toprule
\textbf{Model} & \textbf{Setting} &
\multicolumn{3}{c|}{\textbf{Query~1: Localization}} &
\multicolumn{4}{c}{\textbf{Query~2: Explanation}} \\
\cmidrule(lr){3-5}\cmidrule(lr){6-9}
& &
\textbf{R@3}$\uparrow$ &
\textbf{nDCG}$\uparrow$ &
\textbf{mAP}$\uparrow$ &
\textbf{FieldAcc}$\uparrow$ &
\textbf{CovAvg}$\uparrow$ &
\textbf{ROUGE-L}$\uparrow$ &
\textbf{BERTF1}$\uparrow$ \\
\midrule

LLaVA-OneVision-1.5-8B-Instruct & ZS / Oracle
& 0.192 & 0.190 & 0.133
& 0.532 & 0.267 & 0.035 & 0.372 \\
Qwen2.5-VL-32B-Instruct & ZS / Oracle
& 0.186 & 0.193 & 0.133
& 0.413 & 0.322 & 0.047 & 0.497 \\
Qwen3-VL-8B-Instruct & ZS / Oracle
& 0.241 & 0.244 & 0.180
& 0.565 & 0.700 & 0.073 & \textbf{0.857} \\
InternVL3-78B & ZS / Oracle
& 0.241 & 0.238 & 0.211
& 0.434 & 0.619 & 0.049 & 0.853 \\
\midrule

InternVL3-78B & E2E
& 0.241 & 0.238 & 0.211
& 0.051 & 0.535 & 0.005 & 0.084 \\
Qwen3-VL-8B-Instruct & E2E
& 0.241 & 0.244 & 0.180
& 0.149 & 0.643 & 0.016 & 0.187 \\
GRIDEX & E2E
& \textbf{0.386} & \textbf{0.411} & \textbf{0.333}
& \textbf{0.333} & \textbf{0.884} & \textbf{0.084} & 0.413 \\
\bottomrule
\end{tabular}
\end{adjustbox}
\end{table}

\textbf{Query~1 (Localization).} With 3 random selections out of 16 total regions, the expected nDCG is 0.168. In the localization task, ZS baseline VLMs achieved near-chance nDCG and similar recall rates, suggesting that off-the-shelf models, which are not specifically trained for this purpose, have limited capability in detecting artifacts in spectrograms. Among the baselines, Qwen3-VL-8B-Instruct and InternVL3-78B stood out.

\textbf{Query~2 (Structured Explanations).} 
The ROUGE-L scores are less than 10\%, which is generally low. 
This is expected due to the in-domain terms and naming of audio artifacts in the pseudo-GT explanations, and general-purpose VLMs are not equipped to understand these domain specificities. However, BERTF1 remains in a high range for Qwen3-VL-8B-Instruct and InternVL3-78B, indicating that the explanations are relatively close to $E_i$ in meaning. In particular, Qwen3-VL-8B-Instruct produces the most correct and consistent explanations, scoring highly in coverage, field accuracy, and text similarity. The remaining models perform worse on these criteria, indicating that even when categorical fields can be predicted, the free-form evidence often expands beyond the defined tuple schema and introduces attributes that are not supported by the localized region.

\textbf{Full pipeline.} In the E2E pipeline, Query~2 must explain the regions predicted by Query~1, which may not contain the intended artifact. This propagation of errors causes the performance of Query~2 to deteriorate in both baselines. GRIDEX outperforms baselines in Query~1 across all evaluation metrics. As a result, the E2E Query~2 metrics for GRIDEX are also higher than those of the strong E2E baselines, because Query~2 is conditioned on more frequently correct regions during inference. This behavior is consistent with the Query~2 GRPO schedule, where prediction-conditioned optimization exposes the model to Query~1 region inputs, followed by GT-conditioned refinement. Despite the higher performance, both GRIDEX and the baselines show a significant gap between Query~2 results conditioned on Oracle IDs and those conditioned on predicted IDs, and E2E performance is still greatly affected by the performance of Query~1. Improving localization is the most direct solution to recover explanation quality.

\subsubsection{Per-query effect of staged optimization}
\label{sec:qualitative}
\begin{table}[H]
\centering
\footnotesize
\setlength{\tabcolsep}{3pt}
\renewcommand{\arraystretch}{1.08}
\caption{Per-query evaluation of SFT and subsequent GRPO checkpoints. Query~2 is evaluated with Oracle IDs.}
\label{tab:staged_opt_combined}
\begin{adjustbox}{max width=\columnwidth}
\begin{tabular}{lccc|cccc}
\toprule
\textbf{Stage} &
\multicolumn{3}{c|}{\textbf{Query~1: Localization}} &
\multicolumn{4}{c}{\textbf{Query~2: Explanation}} \\
\cmidrule(lr){2-4}\cmidrule(lr){5-8}
& \textbf{R@3} & \textbf{nDCG} & \textbf{mAP}
& \textbf{FieldAcc} & \textbf{CovAvg} & \textbf{ROUGE-L} & \textbf{BERTF1} \\
\midrule
SFT  & 0.309 & 0.333 & 0.264 & 0.313 & 0.077 & 0.021 & 0.574 \\
GRPO & 0.386 (+0.077) & 0.411 (+0.078) & 0.333 (+0.069) & 0.522 (+0.209) & 0.886 (+0.809) & 0.119 (+0.098) & 0.865 (+0.291) \\
\bottomrule
\end{tabular}
\end{adjustbox}
\end{table}

\textbf{Effect on localization.}
Compared to SFT-1, applying GRPO-1 improves all localization metrics. SFT teaches the model the task format and basic mapping, while GRPO-1 pushes it toward better region selection by optimizing localization rewards directly. The GRPO-1 reward function emphasizes R@3 as the dominant weight because increases in this metric tend to improve ranking quality as well: correct items enter the shortlist more frequently, giving nDCG/mAP a chance to increase. %

\textbf{Effect on structured explanations.}
SFT-2 first establishes and stabilizes the three-tuple output format. At this stage, outputs pass format validation, but the model still struggles with explanation quality. The explanations are not consistently grounded in their declared fields and lean toward free-form generation. Subsequently, GRPO-2 improves structured explanation quality, consistent with the GRPO-2 reward design. %

\textbf{Overall impact.}
Post-SFT GRPO refinement improves both query-specific objectives under per-query evaluation. The E2E results in Table~\ref{tab:baselines_combined} further indicate that the pipeline remains effective when Query~2 is conditioned on Query~1 predictions. %

\subsubsection{Qualitative analysis}
\begin{figure*}[!tbp]
\centering
\includegraphics[width=0.8\textwidth]{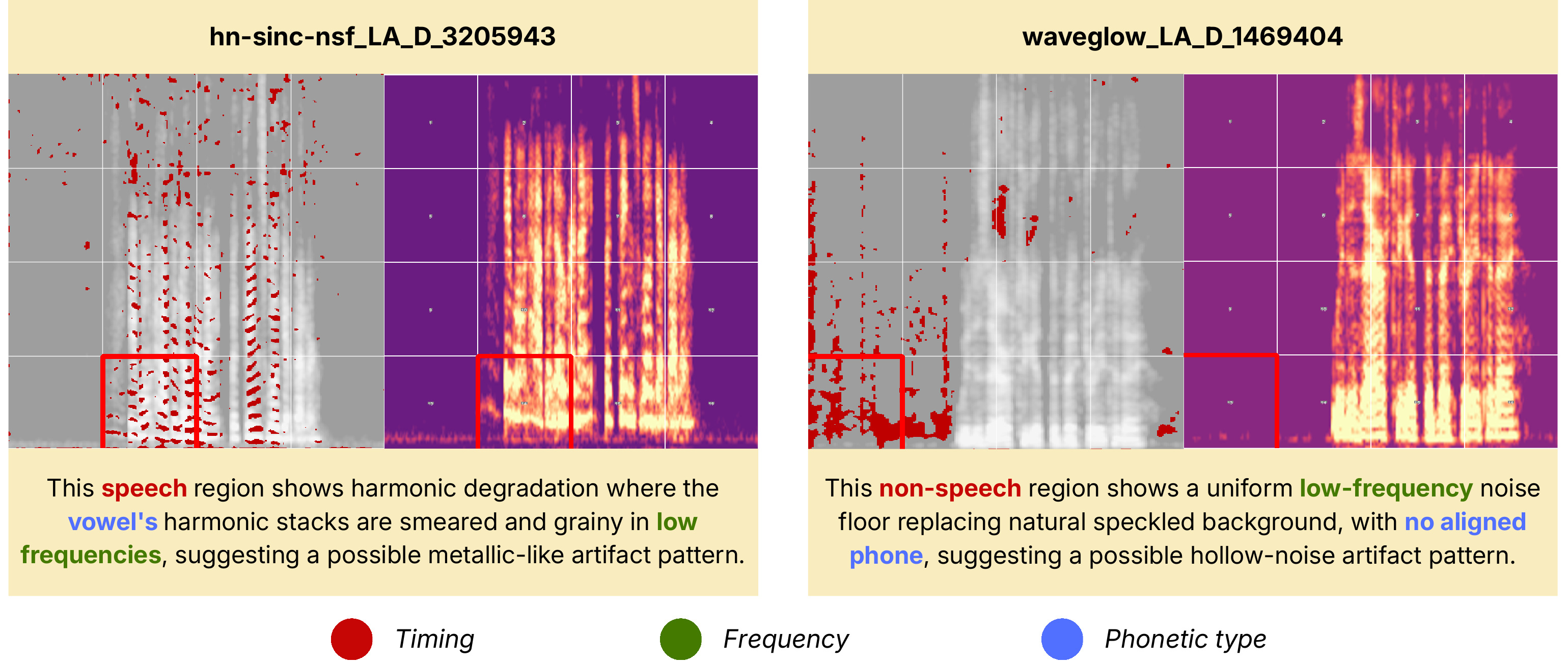}
\caption{Generated explanation samples on most prominent artifacts.}
\label{fig:en_samples}
\end{figure*}

\noindent
During data construction, we use reference cues to guide $E_i$ generation (Sec.~\ref{sec:annotation_pipelines}). These cues are not provided to Query~2 during inference. This design allows generated explanations to align with known artifact patterns without restricting them to a fixed cue list, leaving room for recurring patterns outside the original cues to emerge. Fig.~\ref{fig:en_samples} shows $E_i$ examples where Query~1 localizes artifact regions correctly and Query~2 provides region-grounded interpretations. The model's explanations often align with four categories: 
\begin{itemize}
  \item \textbf{Fogging} \citep{Lee2023PASSIAFPA}.
  Smearing of normally distinct speech patterns (e.g., vowel/consonant structure), commonly reported across synthesis methods.
  \item \textbf{Formant attenuation} \citep{Lee2023PASSIAFPA}.
  Reduction in the definition of vowel formant bands in higher-frequency regions where formant energy becomes weak or partially suppressed.
  \item \textbf{Over-smoothing of stochastic detail.}
  Reduced noise-like microstructure in non-speech and pre-speech frames, a known issue in full-band mel-to-waveform neural vocoders \citep{Jang2021UnivNet}.
  \item \textbf{Imaging} \citep{pons2021upsampling, bak2023avocodo}.
  Spurious high-frequency replicas of lower-frequency structure in vertical stripes, often linked to GAN-based vocoders.
\end{itemize}

Beyond these cue-derived categories, we observe additional recurring artifact patterns among successful predictions:

\begin{itemize}
  \item \textbf{High-frequency discontinuities in between-segment features (7.02\%).}
  The transitions between coarticulatory features appear abrupt in the high-frequency bands.
  \item \textbf{Mid-band formant inconsistencies (5.77\%).}
  Mid-band speech structure deviates from expected resonance patterns.
\end{itemize}

Region-grounded explanations provide an artifact profile that analysts can verify against specific spectrogram regions and compare across cases. Because each claim is tied to a region ID and categorical attributes, the output is easier to audit than unconstrained rationales and can support human-in-the-loop forensic review. The proposed pipeline focuses on common acoustic features, but the framework could be extended to capture more complex properties when the model is given the necessary inputs to detect them.

In contrast, among inaccurate detections, errors show distinct bottlenecks in each of the two queries. Query~1 exhibits hotspot bias, where predictions concentrate on frequent IDs (14--16), producing recurrent top-3 patterns that reduce ranking diversity. Slot-level correctness drops from 29.1\% at rank 1 to 15.1\% and 9.1\% at ranks 2--3, indicating that the model often finds the most salient artifact but misses secondary regions. Query~2 fails primarily due to inaccurate field extraction. While the explanation text is constrained by the reward design, overall explanation quality remains limited by field accuracy. In addition, template drift can produce explanations that are coherent but field-inconsistent, while the absence of reference cues from the Query~2 input at inference limits the use of specific terminology.

\subsection{Ablation Studies}
\label{sec:ablations}
\subsubsection{Granularity levels}
\label{sec:ablation_granularity}
\begin{table}[!htbp]
\centering
\small
\setlength{\tabcolsep}{2.5pt}
\renewcommand{\arraystretch}{1.08}
\begin{tabular}{lccc}
\toprule
\textbf{Setting} & \textbf{R@3}$\uparrow$ & \textbf{nDCG@3}$\uparrow$ & \textbf{mAP}$\uparrow$ \\
\midrule
Grid & 0.309 & 0.333 & 0.264 \\
SLIC & 0.261 (-0.048) & 0.291 (-0.042) & 0.214 (-0.050) \\
\bottomrule
\end{tabular}
\caption{Granularity effect on Query~1.}
\label{tab:ablation_granularity_q1}
\end{table}

Changing the region division from a fixed grid to SLIC superpixels reduces localization performance. As shown in Table~\ref{tab:ablation_granularity_q1}, moving from Grid to SLIC lowers performance in all three metrics. This drop is consistent with prior findings that conventional superpixel methods perform poorly on spectrograms due to weak acoustic boundaries~\citep{aironi2021graph}. SLIC produces irregular regions, which may be less suitable for consistent region-ID selection under difference-mask supervision.

\subsubsection{Hit reward on localization}
\label{sec:ablation_hit_reward}

\begin{table}[!htbp]
\centering
\small
\setlength{\tabcolsep}{2.5pt}
\renewcommand{\arraystretch}{1.08}
\begin{tabular}{lccc}
\toprule
\textbf{Setting} & \textbf{R@3}$\uparrow$ & \textbf{nDCG@3}$\uparrow$ & \textbf{mAP}$\uparrow$ \\
\midrule
GRPO-1 & 0.386 & 0.411 & 0.333 \\
GRPO-1 (w/o \(H\)) & 0.377 (-0.009) & 0.396 (-0.015) & 0.318 (-0.015) \\
\bottomrule
\end{tabular}
\caption{Hit reward ablation on GRPO-1.}
\label{tab:ablation_grpo1_novelty}
\end{table}
 
 Removing $H$, the highest-weighted term in the GRPO-1 reward, degrades localization because the main difficulty in this task is set retrieval. Given that the random-baseline nDCG is around 0.17, meaningful gains require consistently selecting the correct region set, not adjusting ranks among mostly incorrect candidates. This is also reflected by the fact that even if a model always predicts the correct three IDs but in random order, nDCG is around 0.895, implying that most nDCG loss comes from missing the correct IDs, not mis-ordering them. Accordingly, the GRPO-1 reward with $H$ better aligns optimization with this error mode, producing higher metrics than training without it.

\section{Limitations and Future Work}

First, GRIDEX explains samples flagged by an upstream detector (Sec.~\ref{sec:task_overview}). Its outputs should therefore be interpreted as explanations of a given spoof hypothesis, not independent evidence of spoofing. Classification errors and false alarms remain determined by the upstream detector. Future work could incorporate detector signals for detector faithfulness.

Second, Query~2 performance decreases when conditioned on Query~1 predictions compared to Oracle IDs (Table~\ref{tab:baselines_combined}), indicating error propagation from localization to explanation generation. This gap limits E2E reliability and motivates improved localization and greater robustness to imperfect region inputs.

Third, our supervision relies on controlled vocoder data with bona fide--spoof pairs to derive residual region references (Sec.~\ref{sec:data_construction}). The current evaluation covers held-out utterances and speakers within this setting but does not establish generalization to unseen vocoders. Extending GRIDEX to unconstrained TTS and VC settings will require alignment-free supervision and external validation.

Finally, because GRIDEX uses VLM/LLM components and pseudo-reference rationales, generated explanations may contain unsupported acoustic interpretations. Region conditioning, structured fields, verifier checks, and manual correction mitigate this risk but do not replace independent validation. Listener studies, practitioner-facing evaluation, and stronger grounding mechanisms therefore remain important for forensic deployment.

\section{Conclusion}
We propose GRIDEX, a VLM framework for region-grounded forensic explanations of vocoder-generated synthetic speech spectrograms. Given a sample flagged by an upstream detector, GRIDEX performs two sequential tasks: Query~1 identifies the top-$K$ most suspicious regions on a fixed grid, and Query~2 produces a structured per-region explanation tuple that maps each claim to a region ID and its acoustic context. To build supervision data, we construct a region-grounded dataset from difference masks and deterministic metadata, and generate pseudo-reference explanations through an automatic annotation pipeline with manual correction. Our model is trained through staged optimization using task-specific PEFT adapters: GRPO-1 improves set retrieval and ranking, while GRPO-2 improves structured explanations. Evaluations against VLM baselines show that GRIDEX achieves strong E2E pipeline performance. The resulting explanations align with recurring spectrogram artifact categories while also revealing cues beyond the initial reference set. These findings suggest that GRIDEX can support human-in-the-loop forensic review by turning an upstream spoof decision into an auditable, region-grounded rationale. Future work should focus on (i) improving localization diversity and accuracy, (ii) incorporating joint detection--explanation training, and (iii) extending to cross-dataset generalization.

\bibliography{sources}
\appendix
\section{Training Hyperparameters}
We used cosine learning-rate decay and DeepSpeed ZeRO-2 in all stages.
We summarize the SFT hyperparameters in Table~\ref{tab:hparams_sft_combined}
and the GRPO hyperparameters in Table~\ref{tab:hparams_grpo_combined}. %

\begin{table}[h]
\centering
\footnotesize
\setlength{\tabcolsep}{4pt}
\renewcommand{\arraystretch}{1.05}
\begin{adjustbox}{max width=\columnwidth}
\begin{tabular}{@{}lcc@{}}
\toprule
\textbf{Setting} & \textbf{SFT-1} & \textbf{SFT-2} \\
\midrule
Conditioning IDs & GT IDs & GT IDs \\
Initialization & Base model & Base model \\
Precision & bf16 & bf16 \\
LoRA rank / alpha & 8 / 32 & 32 / 64 \\
Per-device batch / grad acc. & 2 / 20 & 2 / 32 \\
Epochs & 3 & 2 \\
Learning rate & $1\times10^{-4}$ & $5\times10^{-5}$ \\
Warmup ratio & 0.03 & 0.03 \\
Max length / pixels & 768 / 524288 & 1408 / 524288 \\
\bottomrule
\end{tabular}
\end{adjustbox}
\caption{SFT hyperparameters for Query~1 and Query~2.}
\label{tab:hparams_sft_combined}
\end{table}

\begin{table}[h]
\centering
\footnotesize
\setlength{\tabcolsep}{4pt}
\renewcommand{\arraystretch}{1.05}
\begin{adjustbox}{max width=\columnwidth}
\begin{tabular}{@{}lccc@{}}
\toprule
\textbf{Setting} & \textbf{GRPO-1} & \textbf{GRPO-2 Phase A} & \textbf{GRPO-2 Phase B} \\
\midrule
Conditioning IDs & GT IDs & Predicted IDs & GT IDs \\
Initialization & SFT-1 checkpoint & SFT-2 checkpoint & Phase-A checkpoint \\
Precision & bf16 & bf16 & bf16 \\
Max length / completion & 768 / 64 & 1408 / 512 & 1024 / 384 \\
Per-device batch / grad acc. & 2 / 16 & 2 / 16 & 4 / 8 \\
Epochs & 2 & 1 & 5 \\
Learning rate & $1\times10^{-5}$ & $1\times10^{-5}$ & $1\times10^{-5}$ \\
Num. generations & 8 & 2 & 8 \\
Reward weights & $[1.0,\;0.3,\;0.1]$ & $[1.0,\;0.5,\;0.1,\;0.1]$ & $[1.0,\;0.5,\;0.1,\;0.1]$ \\
\bottomrule
\end{tabular}
\end{adjustbox}
\caption{GRPO hyperparameters for Query~1 and Query~2.}
\label{tab:hparams_grpo_combined}
\end{table}
\end{document}